\documentclass[%
 preprint,
 superscriptaddress,
 amsmath,amssymb,
 aps,
 prb,
]{revtex4-1}

\usepackage{amsmath}
\usepackage[colorlinks=true,citecolor=blue]{hyperref}
\usepackage{graphicx}
\usepackage{dcolumn}
\usepackage{bm}
\usepackage[dvipsnames]{xcolor}
\usepackage[normalem]{ulem}

\begin{document}

\preprint{APS/123-QED}


\title{Spin relaxation in multilayers with synthetic ferrimagnets}

\author{D.~M.~Polishchuk}
\email{dpol@kth.se.}
\affiliation{Nanostructure Physics, Royal Institute of Technology, Stockholm, Sweden}%
\affiliation{Institute of Magnetism, NAS of Ukraine, Kyiv, Ukraine}

\author{T.~I.~Polek}
\affiliation{Institute of Magnetism, NAS of Ukraine, Kyiv, Ukraine}

\author{A.~Kamra}
\affiliation{Center for Quantum Spintronics, Department of Physics, Norwegian University
of Science and Technology, Trondheim, Norway}

\author{A.~F.~Kravets}
\affiliation{Nanostructure Physics, Royal Institute of Technology, Stockholm, Sweden}
\affiliation{Institute of Magnetism, NAS of Ukraine, Kyiv, Ukraine}

\author{A.~I.~Tovstolytkin}
\affiliation{Institute of Magnetism, NAS of Ukraine, Kyiv, Ukraine}
 
\author{A.~Brataas}%
\affiliation{Center for Quantum Spintronics, Department of Physics, Norwegian University
of Science and Technology, Trondheim, Norway}%

\author{V.~Korenivski}%
\affiliation{Nanostructure Physics, Royal Institute of Technology, Stockholm, Sweden}%

\date{\today}

\begin{abstract}

We demonstrate a strong tunability of the spin-pumping contribution to magnetic damping in a thin-film ferromagnetic free layer interfaced with a synthetic ferrimagnet (SFM), acting as a spin-sink, via a thin Cu-spacer. The effect strongly depends on the magnetic state of the SFM, a trilayer structure composed of two Fe layers coupled via indirect exchange mediated by a Cr spacer. With increasing Cr thickness, the SFM state undergoes a transition from an antiparallel via a non-collinear to a parallel configuration. We can explain the corresponding non-monotonous dependence of spin relaxation in the free layer in terms of a modulation of the longitudinal spin transport as well as relaxation of the transverse angular momentum in the SFM. The results should be useful for designing high-speed spintronic devices where tunability of spin relaxation is advantageous. 

\end{abstract}

\maketitle



\section{Introduction}

{Spin pumping, when a precessing ferromagnetic layer pumps spin-polarized electrons into an adjacent nonmagnetic layer \cite{Tserkovnyak2002}, leads to a number of related effects in magnetic multilayers and can be used for tuning spin relaxation, important for magnetization switching in spintronic devices \cite{Brataas2002,Ando2011}. According to its phenomenology, spin-pumping is a reciprocal effect to that of spin-transfer torque \cite{Tserkovnyak2005,SpinCurrent}. A ferromagnetic layer, whose precession can be activated via ferromagnetic resonance (FMR), emits spin-polarized electrons, thereby losing spin angular momentum, which enhances its effective Gilbert damping. In magnetic multilayers containing non-magnetic spacers (N) interfacing multiple ferromagnetic layers, spin pumping induces nonequilibrium spin currents in N, which in turn affect the magnetic dynamics of the spin-emitting layer as well as the ferromagnetic layers interfaced to it. If the neighboring magnetic layers, having the same FMR conditions, are simultaneously excited, spin pumping can lead to a syncronization of the magnetization precession in the multilayer, producing the interlayer coupling known as the dynamic exchange \cite{Heinrich2003}. Dynamic exchange and its related effects have been successfully described by combining the magnetoelectronic circuit theory \cite{Brataas2000} with that for adiabatic spin-pumping \cite{Tserkovnyak2002,Tserkovnyak2002a,Tserkovnyak2005}.

The spin-pumping contribution to magnetization damping can be used as an effective microscopic probe of the interface properties in magnetic multilayered systems \cite{Ghosh2012,Merodio2014} due to its dynamic nature and long-range character (the order of the spin-flip length, $\lambda$). Spin-pumping has been used to determine the spin-absorption and spin-mixing conductivities in thin films \cite{Mizukami2001,Foros2005}. In multilayers, this dynamic spin-dependent probe can sense the spin penetration depth in ultrathin ferromagnetic layers \cite{Ghosh2012}. Moreover, the spin-pumping effect is widely-used for characterizing the spin penetration depth and spin relaxation mechanisms in antiferromagnets \cite{Merodio2014}, which is important for antiferromagnetic spintronics \cite{Jungwirth2016} since the more standard experimental tools, such as magnetoresistance measurements~\cite{Bass2007}, are not easily applicable to studying spin transport in antiferromagnets. 

When one ferromagnetic layer in a multilayer is excited via FMR, the other, \emph{static} ferromagnetic layers usually play the role of passive spin sinks for the spin-pumped current out of the layer under FMR. The efficiency of the passive layers' spin-scattering and spin-absorption can be defined by a number of parameters such as the layers' thickness, type of magnetic order (ferro-, antiferro-, para-magnetic), interface characteristics. Another important parameter is the orientation of the magnetization with respect to the resonating layer (F). The parallel/antiparallel and non-collinear magnetization configurations can lead to different distributions of the spin currents in the interfacing nonmagnetic layers (N). As schematically shown in Fig.~\ref{fig_1}(a)-(b), the magnetization orientation in the non-resonant (static) layer changes the spin accumulation in N, which in turn affects the spin dynamics in F. The antiparallel/parallel configuration [Fig.~\ref{fig_1}(a)] leads mainly to enhanced spin scattering of the spin-pumped current, $\mathbf{I}_\text{s}^\text{pump}$, at the interfaces and in the bulk of the static layers [SFM in Fig.~\ref{fig_1}(a)-(b)]. The non-collinear alignment [Fig.~\ref{fig_1}(b)] additionally contributes to spin-absorption of the transverse component of $\mathbf{I}_\text{s}^\text{pump}$ via the same mechanism as that of spin-transfer-torque~\cite{Slonczewski1996}. This contribution depends on the magnetization misalignment angle $\Delta \phi$ between the F/N and N/SFM interfaces as $\sin^2 \Delta \phi$, and remains effective even for thin layers (thickness much smaller than $\lambda$). 

\begin{figure}[b]
\includegraphics[width=1\linewidth]{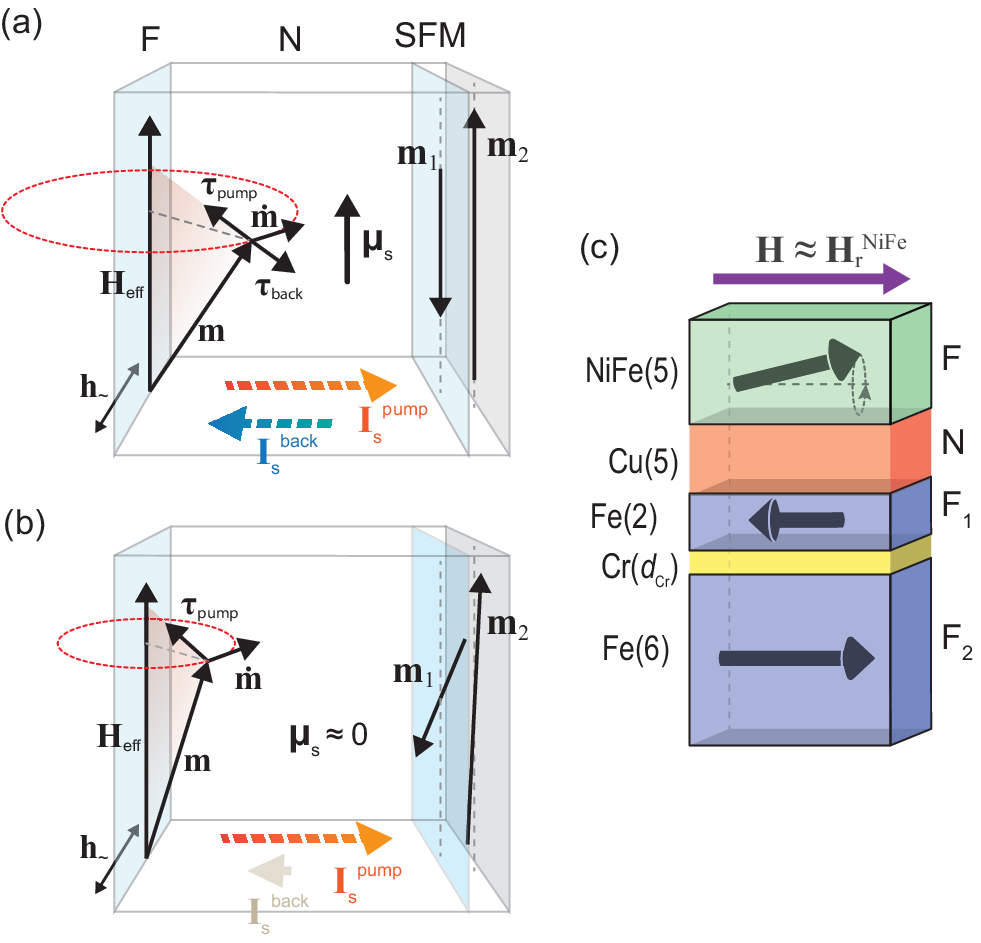}
\caption{Illustration of macrospin dynamics of ferromagnetic layer F and spin-current distribution in nonmagnetic layer N interfacing F and synthetic ferrimagnetic layer SFM for (a) antiparallel and (b) non-collinear alignment of F and SFM magnetic moments $\mathbf{m}_1$ and $\mathbf{m}_2$. Magnetization dynamics of F, excited by FMR microwave field $\mathbf{h}_\sim$, strongly depends on spin accumulation $\boldmath{\mu}_\text{s}$ in spacer N, created by spin currents $\mathbf{I}_\text{s}^\text{pump}$ and $\mathbf{I}_\text{s}^\text{back}$. (c) Fabricated multilayer layout and material composition, with layer thicknesses given in parenthesis in nanometers ($d_\text{Cr} =$ 1.2–5~nm).}
\label{fig_1}
\end{figure}

The spin-pumping related magnetisation damping for the non-collinear magnetization configuration has been investigated theoretically \cite{Heinrich2003,SpinCurrent,Taniguchi2007,Chiba2015} and to some extent experimentally \cite{Mosendz2008,Demirtas2011}. However, no comprehensive experiment has been reported on how spin pumping, relaxation, and precession vary versus the respective changes in the mutual magnetization orientation. The key issue, at least for the standard FMR measurement layout discussed in the above studies, has likely been the difficulty to realize a well-controlled non-collinear and/or antiparallel state. Namely, exchange biasing by an antiferromagnet used to create unidirectional anisotropy in one of the ferromagnetic layers \cite{Nogues2005} (a few hundred Oe at room temperature) or intrinsic magnetic anisotropy \cite{Demirtas2011} are usually too weak to ensure anti-/non-collinear alignment for typical FMR biasing fields (kOe range for transition-metal ferromagnets). Therefore, typical FMR experiments study predominantly the parallel magnetization state, where the the spin-pumping effect results in enhanced magnetization damping \cite{Timopheev2014,Kravets2016}. 

In this work, we implement a novel multilayer design, depicted in Fig.~\ref{fig_1}, which enables us to systematically study the spin-pumping contribution to the magnetization dynamics for parallel, antiparallel, as well as non-collinear mutual alignment of the resonating and static magnetic layers. The key element of the design is a synthetic ferrimagnet (SFM) trilayer Fe(2)/Cr($d_\text{Cr}$)/Fe(6) (F$_1$/Cr/F$_2$), where the F$_1$ and F$_2$ layers are of different thickness, coupled antiferromagnetically by strong Ruderman-Kittel-Kasuya-Yosida (RKKY) exchange \cite{Gruenberg1987} through the thin Cr spacer. The free layer [permalloy, NiFe(5)] is separated from the SFM by a Cu spacer (5 nm thick, no static interlayer coupling). In an external field $\mathbf{H}$, the F$_2$ magnetic moment is fixed along $\mathbf{H}$ due to its larger thickness, while the smaller F$_1$ moment can be parallel or antiparallel to it depending on the strength of $H$ (whether or not $H$ exceeds the RKKY-exchange field). The RKKY-exchange bias of F$_1$ can exceed 2~kOe, enabling the antiparallel mutual orientation of F$_1$ and F magnetic moments at the NiFe (F) resonance field of the in-plane FMR configuration ($H_\text{r}^\text{NiFe} \approx$ 1.3~kOe at 9.9~GHz). By changing $d_\text{Cr}$, we control the strength of the RKKY coupling between F$_1$ and F$_2$ and, thereby, vary the mutual orientation of F$_1$ and F between antiparallel, parallel, and non-collinear at $H \approx H_\text{r}^\text{NiFe}$.

\section{Experimental results}

\subsection{Samples and measurement methods}

The main series of multilayered samples had the composition of Ni$_{80}$Fe$_{20}$(5)/Cu(5)/Fe(2)/ Cr($d_\text{Cr}$)/Fe(6), where $d_\text{Cr} =$ 1.3, 1.5, 1.7, 2, and 5~nm. The thicknesses in parentheses are in `nm'. The last three layers in these structures constitute the  SFM, which was investigated independently via a separate reference series comprised of Fe(2)/Cr($d_\text{Cr}$)/Fe(6) trilayers, with the same sequence of $d_\text{Cr}$. The reference single-layer sample, against which the magnetic damping in the samples of the main series was compared, was a 5-nm-thick Ni$_{80}$Fe$_{20}$ film deposited onto a Cr(5)/Cu(5) buffer bi-layer. The multilayers were deposited by dc-magnetron sputtering at room temperature onto Ar pre-etched un-doped Si (100) substrates. Ni$_{80}$Fe$_{20}$ films (hereafter referred to as NiFe) were deposited using co-sputtering from separate Ni and Fe targets. The composition of the NiFe layers was controlled by setting the corresponding deposition rates of the individual Ni and Fe components, with relevant calibrations obtained by subsequent thickness profilometry. The in-plane magnetization measurements were performed using a vibrating-sample magnetometer (VSM by Lakeshore Inc.). The FMR measurements were performed using an X-band Bruker ELEXYS E500 spectrometer, with the external magnetic field $\mathbf{H}$ applied in the film plane and perpendicular to the rf field $\mathbf{h}_\sim$. The FMR spectra were measured using two configurations: (i) $\mathbf{h}_\sim$ was oriented in the film plane ($\mathbf{h}_\sim \perp \mathbf{n}$) and (ii) perpendicular to the sample surface ($\mathbf{h}_\sim \parallel \mathbf{n}$). ($\mathbf{n}$ is the normal to the sample surface.) All measurements were performed at room temperature.

\subsection{Static magnetization curves}

The in-plane magnetization curves are obtained by recording the total magnetic moment $m$ of the main-series samples versus a varying applied magnetic field $H$. The ensuing curves, normalized by saturation magnetic moment $m_\mathrm{s}$ and depicted in Fig.~\ref{fig_2}, exhibit three distinct contributions from the three magnetic layers. Sweeping the magnetic field down from the saturation, the magnetic moment of the thin Fe(2) layer switches first due to the antiferromagnetic RKKY-exchange coupling to the thick Fe(6) layer. The latter, being the thickest, aligns with the applied magnetic field since that corresponds to the lowest Zeeman energy for the system. On reversing the applied field, the soft NiFe layer switches first, in a relatively low field, after which a simultaneous switching of the Fe(6) and Fe(2) layers occurs. The saturation field ($H_\text{s}$) is sensitive to the thickness of the Cr spacer ($d_\text{Cr}$). The strong decrease in $H_\mathrm{s}$ for samples with larger $d_\text{Cr}$ reflects the weakening RKKY exchange coupling between Fe(6) and Fe(2), which can be considered effectively absent at and above $d_\text{Cr} =$ 5~nm \cite{Polishchuk2017}.

\begin{figure}
\includegraphics[width=1\linewidth]{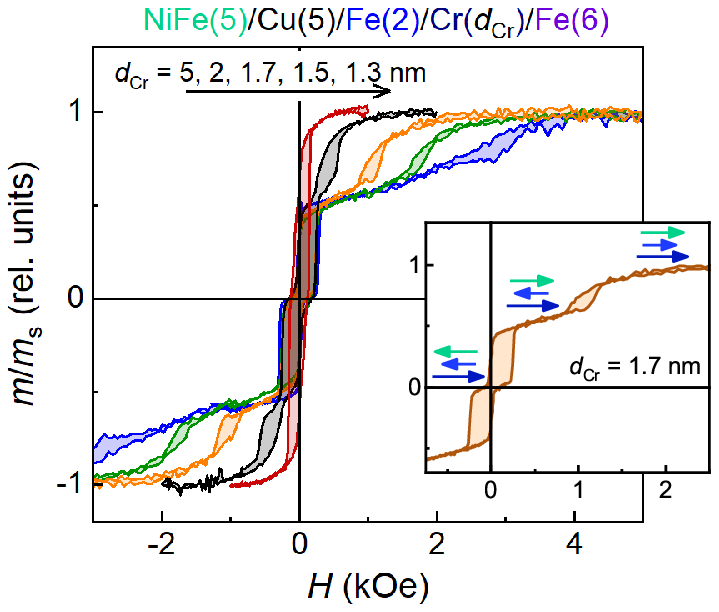}
\caption{ In-plane magnetization loops of main sample series as a function of Cr thickness in SFM Fe(2)/Cr($d_\text{Cr}$)/Fe(6). All curves are normalized by saturation magnetization $m_\mathrm{s}$. Saturation field $H_\mathrm{s}$ increases with decreasing $d_\text{Cr}$. Inset shows selected loop at $d_\text{Cr} =$ 1.7~nm. Arrows represent relative orientation of magnetic moments of NiFe, Fe(2) and Fe(6) layers in different magnetization-vs-field regions.}
\label{fig_2}
\end{figure}

\subsection{Magnetic resonance}

All in-plane FMR spectra for the main sample series for both measurement configurations $\mathbf{h}_\sim \perp \mathbf{n}$ [Fig.~\ref{fig_3}(a)) and $\mathbf{h}_\sim \parallel \mathbf{n}$ (Fig.~\ref{fig_3}(b)] reveal a distinct, high intensity line corresponding to the soft NiFe layer. With increasing the Cr thickness, an additional signal rises in intensity [marked L$_\text{A}$ in panels (a) and (b)] and has the same position as the resonance line of the reference 6-nm thick single-layer Fe film (not shown). The second additional resonance line (L$_\text{N}$) is  highly intensive only in the configuration $\mathbf{h}_\sim \parallel \mathbf{n}$ and exhibits a strong dependence of its position (resonance field) on $d_\text{Cr}$ [Fig.~\ref{fig_3}(b)]. These two additional lines can be assigned to the Fe(2)/Cr/Fe(6) SFM trilayer. Resonance line L$_\text{A}$ has a considerably reduced intensity at smaller Cr thicknesses ($d_\text{Cr} \leq $ 1.7~nm) due to the strong RKKY coupling in the Fe(2)/Cr/Fe(6) trilayer \cite{Zhang1994}, and is not distinguishable in the spectra containing the much stronger NiFe line. The observed behavior of L$_\text{N}$ requires a more detailed discussion presented below.

\begin{figure}
\includegraphics[width=1\linewidth]{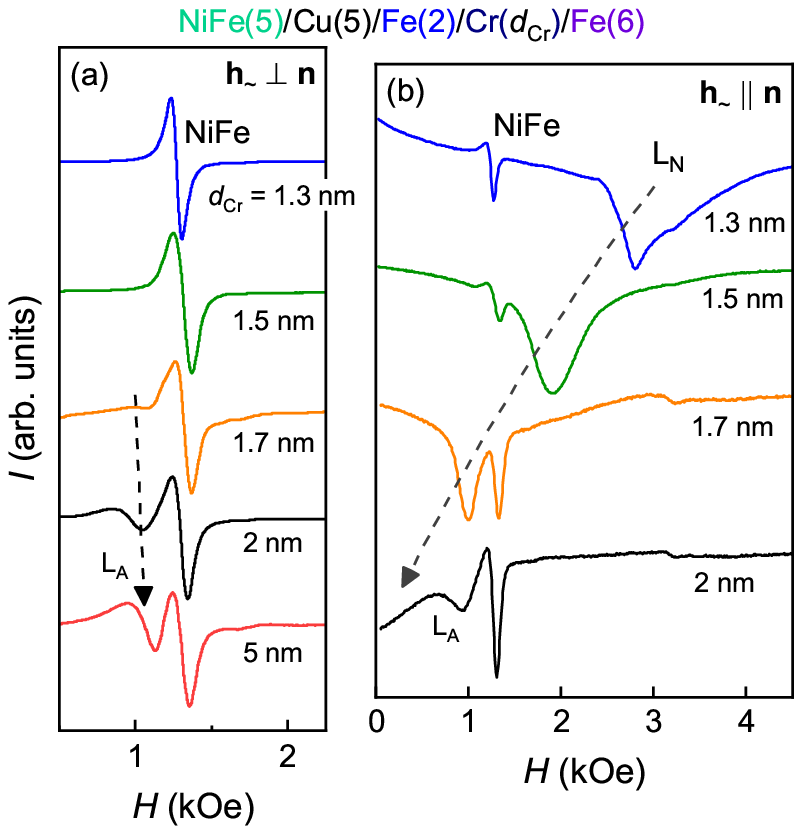}
\caption{ In-plane FMR spectra of main sample series as a function of $d_\text{Cr}$ for measurement configurations with (a) $\mathbf{h}_\sim \perp \mathbf{n}$ and (b) $\mathbf{h}_\sim \parallel \mathbf{n}$. Labels NiFe, L$_\text{A}$ and L$_\text{N}$ mark, respectively, resonance lines for NiFe layer and two resonance modes emerging from Fe(2)/Cr($d_\text{Cr}$)/Fe(6) trilayer. Dashed arrows trace resonance line L$_\text{A}$ in panel (a) and L$_\text{N}$ in panel (b).}
\label{fig_3}
\end{figure}

In order to better understand the observed resonance spectra, we now turn our attention to the reference SFM samples (Fig.~\ref{fig_4}). The signal intensity from the trilayer with the smallest in the series $d_\text{Cr} =$ 1.3~nm is very weak, about the same level as the background (BG) signal from the paramagnetic impurities in the sample holder [see the BG-marked signal line at $\approx$~3.2~kOe in Fig.~\ref{fig_4}(a)], as well as comparable to the noise floor of the measurement (the resonance signal at $\approx$~5.6~kOe). The same background signal can be used as the intensity calibration signal for the spectra of the reference SFM-trilayer samples, which yields clearly stronger resonance lines for $d_\text{Cr} \geq$ 2~nm shown in Fig.~\ref{fig_4}(b).

The reference SFM samples reveal three signals, namely, L$_\text{A}$, L$_\text{O}$ and L$_\text{N}$ in Fig.~\ref{fig_4}, which exhibit different behavior as a function of the Cr thickness. The position of L$_\text{A}$ is insensitive to the changes in $d_\text{Cr}$ and coincides with that of the resonance line of the aforementioned reference 6-nm Fe film. In contrast, the location of L$_\text{O}$ strongly depends on $d_\text{Cr}$ and approaches the position of L$_\text{A}$ from the high-field side as $d_\text{Cr}$ is increased. The difference in resonance fields between L$_\text{A}$ and L$_\text{O}$ is roughly the saturation field on the magnetization curves in Fig.~\ref{fig_2}. It should be noted that the intensity of L$_\text{O}$ is much weaker (an order of magnitude weaker) compared with that for the L$_\text{A}$ line. These resonance lines can be interpreted as the collective modes of the SFM, originating from the RKKY-exchange coupling between the Fe(2) and Fe(6) layers. 

\begin{figure}
\includegraphics[width=1\linewidth]{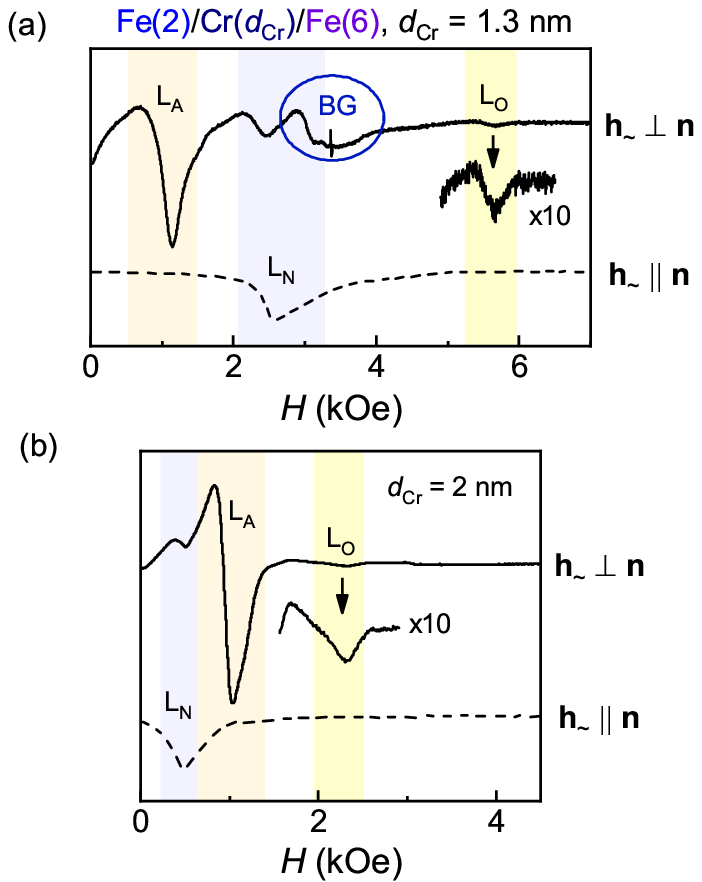}
\caption{ FMR spectra for selected reference samples Fe(2)/Cr($d_\text{Cr}$)/Fe(6) with $d_\text{Cr} =$ 1.3~nm (a) and 2.0~nm (b), obtained in configuration $\mathbf{h}_\sim \perp \mathbf{n}$ (solid lines) and $\mathbf{h}_\sim \parallel \mathbf{n}$ (dashed lines). L$_\text{A}$, L$_\text{O}$, L$_\text{N}$, and BG denote observed three resonance lines from SFM trilayer and background signal, respectively.}
\label{fig_4}
\end{figure}

The position of L$_\text{N}$ is also sensitive to changes in $d_\text{Cr}$ [Fig.~\ref{fig_3}(b)]. However, unlike L$_\text{O}$,  L$_\text{N}$ is observed on both sides of the L$_\text{A}$ resonance field (Fig.~\ref{fig_4}) and its position tracks the switching field of Fe(2) determined from the magnetometry data (Fig.~\ref{fig_2}). Additionally, L$_\text{N}$ has a much higher intensity compared with the other signals for the measurement configuration $\mathbf{h}_\sim \parallel \mathbf{n}$ (dashed lines in Fig.~\ref{fig_4}). We attribute this resonance line to the magnetization dynamics of the unsaturated SFM, as discussed below.

\subsection{Linewidth and damping}

The resonance linewidth ($\Delta H$) for the NiFe layer was extracted from the spectra obtained in the two measurement configurations, $\mathbf{h}_\sim \perp \mathbf{n}$ [Fig.~\ref{fig_3}(a)] and $\mathbf{h}_\sim \parallel \mathbf{n}$ [Fig.~\ref{fig_3}(b)]. $\Delta H$ was then converted into the damping parameter $\alpha$ using $\alpha = 2\Delta H/3\gamma$. The key result of this work is the dependence of $\alpha$ on the Cr thickness shown in Fig.~\ref{fig_5}. The top panel in Fig.~\ref{fig_5} illustrates the three SFM states: (1) parallel orientation of all magnetic moments in the multilayer for $d_\text{Cr} \geq$ 2~nm; (2) non-collinear orientation of the Fe(2) and Fe(6) moments for 1.5~$\leq d_\text{Cr} <$ 2~nm, with the NiFe(5) moment along the applied field; and (3) antiparallel alignment of the Fe(2) moment to those of Fe(6) as well as NiFe(5) for $d_\text{Cr} <$ 1.5~nm.

\begin{figure}
	\includegraphics[width=0.9\linewidth]{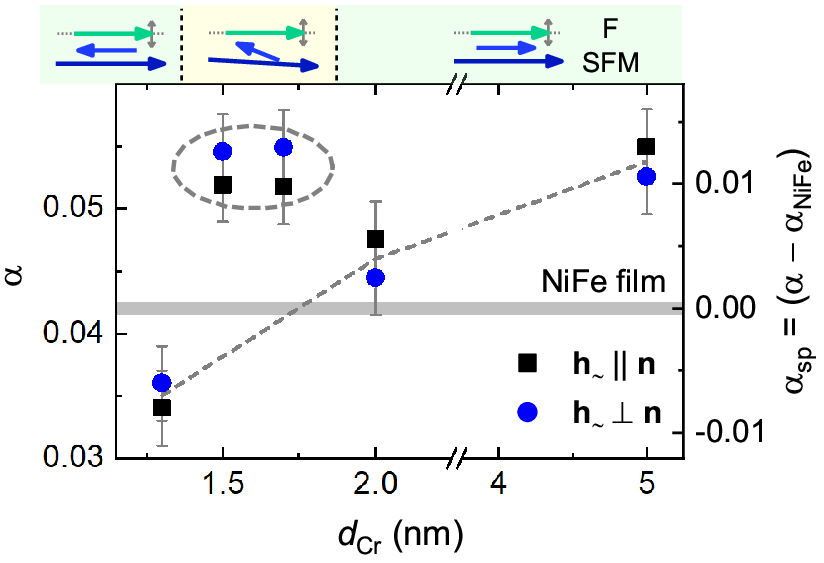}
	\caption{ Damping parameter $\alpha$ of NiFe free layer as a function of SFM-spacer thickness $d_\text{Cr}$; $\alpha_\text{NiFe}$ is damping of single NiFe reference film (grey horizontal line). Top panel illustrates corresponding configurations of resonating F and static SFM, as $d_\text{Cr}$ is varied.}
	\label{fig_5}
\end{figure}

As detailed in the following section, we interpret this dependence of the damping parameter $\alpha$ in NiFe on the Cr thickness in terms of an effective spin relaxation length of the SFM, $\lambda_\mathrm{SFM}$, for the anti-/collinear alignment, whereas the nonmonotonous enhancement of $\alpha$ for the non-collinear configuration is explained within the mechanism of spin-transfer-torque absorption at the N/SFM interface. 

When NiFe is irradiated with a microwave field, a precession of its magnetization about the equilibrium orientation is excited. This leads to a spin-pumping current~\cite{Tserkovnyak2002,Tserkovnyak2005} injected into the adjacent Cu spacer, which transmits it to the SFM without a significant reduction. The SFM absorbs part of the spin current, which is dissipated via spin relaxation, parametrized by $\lambda_\mathrm{SFM}$. The remaining current is reflected and constitutes the so-called `backflow'~\cite{Tserkovnyak2005}. Hence, the net spin current lost by NiFe is given by the difference between the spin-pumping and backflow currents, which in turn is roughly equal to the current absorbed and relaxed by the SFM. The spin current lost by NiFe accounts for the additional damping in the magnetization dynamics~\cite{Tserkovnyak2002}. Thus, the enhancement in damping of the NiFe magnetization precession is larger when the spin relaxation in SFM is stronger. Our results can then be understood as due to an increase in spin relaxation in the SFM for increasing Cr thickness. While comparing the parallel and antiparallel states, an additional factor may come into play -- the spin dependent conductances of the SFM should be different for these two states, similar to the mechanism of the giant magnetoresistance effect~\cite{Valet1993}. In the field region, where the SFM is transiting from the parallel to the antiparallel state, we observe hysteresis in the static magnetization curves (Fig.~\ref{fig_2}) as well as the L$_\mathrm{N}$ mode in the FMR spectra [Fig.~\ref{fig_3}(b)]. This behavior is naturally associated with the unsaturated state of the SFM, having a non-collinear alignment of the interface magnetization with that of the free NiFe layer. If the resonance field of NiFe is within the field interval of the SFM's non-collinear regime, the transverse component of the spin-pumped current is effectively absorbed at the Cu/SFM interface via spin-transfer-torques, thus, enhancing the magnetic damping in NiFe.

\section{Theoretical modeling}

\subsection{Static and dynamic properties of SFM}

SFM is the key  element in our multilayer system and is composed of two ferromagnetic layers of different thicknesses ($t_1 < t_2$) coupled by RKKY-exchange through a thin nonmagnetic spacer. The ferromagnetic layers of the SFM are taken to have the same saturation magnetization, $|\mathbf{M}_1| = |\mathbf{M}_2| = M$, as experimentally both layers are made of the same material (Fe). The magnetic free energy of the system consists of several contributions. Our prior FMR study revealed no in-plane magnetic anisotropy in the polycrystalline Fe films, so we take into account only the perpendicular anisotropy contributions in the form of the effective anisotropy field $\mathbf{H}_{\perp,i}$ ($i =$ 1, 2). This effective perpendicular anisotropy includes the thin-film demagnetizing field, $–4\pi M$, as well as possible contributions from crystallographic, strain, or surface anisotropy -- individually indistinguishable in an FMR measurement. The magnetic free energy per unit surface area is then

\begin{equation}
W = -\sum_{i=1}^2 t_i \mathbf{M}_i (\mathbf{H} + 1/2 \mathbf{H}_{\perp,i}) - J (\mathbf{M}_1 \cdot \mathbf{M}_2)/M^2,
\label{Fen}
\end{equation}

\noindent where the first and second terms under the sum are, respectively, the Zeeman energy in an external field $\mathbf{H}$ and the magnetic anisotropy of the $i$-th ferromagnetic layer. The RKKY bilinear exchange constant per unit area $J$ can be expressed through an effective exchange field $H_\text{ex}$ averaged over the layer. Experimentally, $H_\text{ex}$ equals the saturation field of the magnetization curve and relates to $J$ as \cite{Zhang1994}

\begin{equation}
J = M H_\text{ex} \frac{t_1 t_2}{t_1 + t_2}.
\label{Jrkky}
\end{equation}

Fig.~\ref{fig_6}(a) compares the experimental magnetization curves for the main-series samples to those for the Fe(2)/Cr($d_\text{Cr}$)/Fe(6) trilayers simulated by minimizing the free energy \eqref{Fen}. For a proper comparison, the calculated curves were scaled by the magnetic moment of the NiFe(5) layer and fitted to the experimental data shown as filled grey loops in Fig.~\ref{fig_6}(a). This approach yields the field dependence of the total magnetic moment of the SFM and the equilibrium orientations of the Fe(2) and Fe(6) layers, angles $\phi_1$ and $\phi_2$, respectively [Fig.~\ref{fig_6}(b)]. Inset to Fig.~\ref{fig_6}(b) illustrates the mutual orientation of $\mathbf{H}$ and the SFM magnetic moments $\mathbf{m}_1$ and $\mathbf{m}_2$ lying in the film plane. The exchange constant $J$ can then be obtained for each multilayer composition. Using these $J$ values, the respective exchange fields $H_\text{ex}$ were calculated using \eqref{Jrkky} and found to be in good agreement with those measured in the experiment.

\begin{figure}
\includegraphics[width=0.9\linewidth]{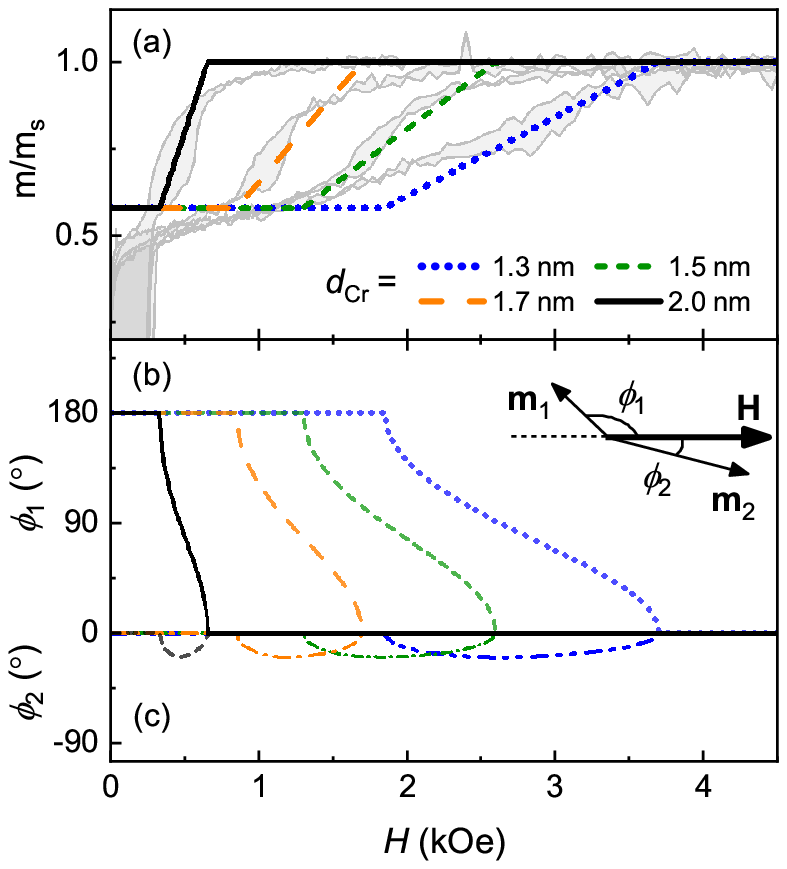}
\caption{ (a) Simulated magnetization curves for SFM Fe(2)/Cr($d_\text{Cr}$)/Fe(6), scaled to the respective data for the main series (filled grey loops from Fig.~\ref{fig_2}). (b),(c) Calculated field dependence of equilibrium angles $\phi_1$ and $\phi_2$ for Fe(2) and Fe(6) magnetic moments, respectively. Inset in panel (b) illustrates mutual orientation of $\mathbf{H}$ and SFM's magnetic moments $\mathbf{m}_1$ and $\mathbf{m}_2$.}
\label{fig_6}
\end{figure}

The above model calculations yield three distinct magnetic states of the SFM trilayer, with the interstate transitions controlled by sweeping the external magnetic field. The high-field and low-field regions exhibit, respectively, parallel and antiparallel orientations of the Fe(2) and Fe(6) magnetic moments. An additional, non-collinear alignment of the SFM's two magnetic moments is found in the intermediate field range and is clearly seen in Fig.~\ref{fig_6}(b). The above analysis shows that at the resonance field of the NiFe(5) layer in the samples of the main series, $H_\text{r}^\text{NiFe} \approx$ 1.3~kOe, the magnetic configuration of the SFM layer can be parallel ($d_\text{Cr} =$ 2, 5~nm), non-collinear ($d_\text{Cr} =$ 1.7, 1.5~nm), or antiparallel ($d_\text{Cr} =$ 1.3~nm).

We next clarify the origin of the three resonance lines, L$_\text{A}$, L$_\text{O}$ and L$_\text{N}$, observed for the exchange coupled Fe(2)/Cr/Fe(6) trilayers, as shown in Fig.~\ref{fig_4}. The dispersion relation for the exchange coupled trilayer, which can be obtained using the magnetic free energy of the system \eqref{Fen}, yields only two resonance modes at any given field \cite{Zhang1994}. One mode (acoustic) corresponds to the two magnetization vectors resonating in-phase and, hence, the dispersion relation of this mode is degenerate with that of a single-layer system. In our case, the acoustic mode is attributed to the L$_\text{A}$ line, the position (resonance field) of which is insensitive to the changes in $d_\text{Cr}$ and coincides with the resonance line of the reference 6-nm Fe film. The optical mode corresponds to the magnetization vectors resonating out-of-phase. Due to the extra exchange coupling \eqref{Jrkky}, for constant frequency, the resonance field of the optical mode is shifted with respect to that of the acoustic mode by $H_\text{ex}$. When the external field $\mathbf{H}$ is applied in the film plane, the optical mode is observed at a higher field than the acoustic mode for the case of antiferromagnetic coupling ($J < 0$) and at a lower field for the case of ferromagnetic coupling ($J > 0$). Since the magnetometry measurements clearly show antiferromagnetic coupling of Fe(2) and Fe(6) in our SFM, one would expect the optical mode always to be at higher fields as compared to the position of L$_\text{A}$. In our case, the optical mode is attributed to L$_\text{O}$, the location of which is strongly dependent on $d_\text{Cr}$ and approaches the position of L$_\text{A}$ from the high-field side as $d_\text{Cr}$ becomes large, with the difference in position of about $H_\text{ex}$ for each sample.

The third resonance line L$_\text{N}$ is sensitive to changes in $d_\text{Cr}$ and is highly intensive in the FMR spectra measured with $\mathbf{h}_\sim \parallel \mathbf{n}$, where the other signals from the SFM trilayer (acoustic and optical modes) are significantly suppressed in intensity [Fig.~\ref{fig_3}(b) and dashed curve in Fig.~\ref{fig_4}]. The position of the L$_\text{N}$ signal shown in Fig.~\ref{fig_3}(b) follows the orientational transition of the Fe(2) layer in SFM, analysed in Fig.~\ref{fig_6}(b). The observed hysteresis and some transition-smearing in the experimental magnetization curves [Fig.~\ref{fig_6}(a)] as well as the broadening of the L$_\text{N}$ signal linewidth lead to the conclusion that the orientational transition is of inhomogeneous character, with potentially domain-like intermediate states. The fact that the deposited films have polycrystalline nature suggests a presence of imperfections at the crystallites' boundaries that can act as nucleation centers for domains or at least local spin perturbations. We thus speculate that the L$_\text{N}$ line is due to an unsaturated resonance, reflecting an inhomogeneous, non-collinear magnetic state of SFM.

\subsection{Spin pumping mediated damping}

In this section, we discuss a qualitative theory of spin pumping mediated damping in the multilayered structure of interest. The FMR driven F layer pumps spin current into N~\cite{Tserkovnyak2002}. A part of it is absorbed and dissipated by the SFM, while the remaining current is returned to F as backflow (Fig.~\ref{fig_1}). It is the spin current absorbed by the SFM which constitutes the net loss of spin by the F layer and thus leads to an enhanced damping. For simplicity, we model the SFM as a single ferromagnetic layer with its spin relaxation length depending on the magnetic states of the two constituent Fe layers. Thus our multilayer structure may be regarded as an effective trilayer F/N/SFM within a simplified picture.

In the limit of negligible spin-flip scattering in the system, F/N/F structures are predicted to exhibit highest Gilbert damping for collinear magnetizations in the two ferromagnets~\cite{Tserkovnyak2005}. Furthermore, in this limit, the spin pumping mediated enhancement in Gilbert damping is independent of the spin-sink layer thickness. In arriving at these conclusions~\cite{Tserkovnyak2005}, it is additionally assumed that the FMR condition for the driven magnetic layer F is well separated from that for the spin-sink layer F, such that the latter may be treated as static, thereby circumventing any effects of dynamical exchange coupling~\cite{Heinrich2003}. 

Based on our experimental observations, we infer that dynamical exchange may be disregarded. However, a strong spin relaxation in the SFM layer leads to a behavior different from the above stated expectations, which are based on the dominance of transverse spin absorption. Thus, with the aim of achieving a qualitative understanding, we assume that the dominant contribution to the spin current absorbed by the SFM comes from the longitudinal spin transport and relaxation. A detailed model including the transverse contributions as well as anisotropies shall be discussed elsewhere.

The effective spin mixing conductance, that determines the spin current lost by F after accounting for the backflow, is given by:

\begin{align}\label{geff}
\frac{1}{g_{\mathrm{eff}}} = & \frac{1}{g_{\mathrm{F}}} + \frac{1}{g_{\mathrm{SFM}}},
\end{align}

\noindent where $g_{\mathrm{F}}$ is the real part of the spin mixing conductance of the F/N interface and $g_{\mathrm{SFM}}$ is the longitudinal spin conductance of the SFM layer, modeled as a single ferromagnetic layer as discussed above. We have disregarded the spin relaxation in the N layer and the typically small imaginary part of the spin mixing conductance, for simplicity. $g_{\mathrm{SFM}}$ is evaluated as $4 \pi$ times the ratio of the longitudinal spin current injected into the SFM layer by a longitudinal spin chemical potential at its interface with N. To this end, we solve the diffusion equation in the SFM, which is treated as a single magnetic layer with effective thickness $L_{\mathrm{SFM}}$, spin relaxation length $\lambda_{\mathrm{SFM}}$, and spin dependent conductivities $\sigma_{\uparrow,\downarrow}$, while imposing the boundary condition of vanishing spin current at the far end of the SFM. The longitudinal spin conductance is obtained as~\cite{Fabian2010}:

\begin{align}\label{gb}
g_{\mathrm{SFM}} = & \frac{4 \pi \hbar S \sigma_{\uparrow} \sigma_{\downarrow}}{e^2 \left( \sigma_{\uparrow} + \sigma_{\downarrow} \right) \lambda_{\mathrm{SFM}}} \tanh \left( \frac{L_{\mathrm{SFM}}}{\lambda_{\mathrm{SFM}}} \right),
\end{align}

\noindent where $S$ is the interfacial area. From Eqs. (\ref{geff}) and (\ref{gb}), we note that a larger enhancement in Gilbert damping results from an increasing $g_\mathrm{SFM}$.

\section{Discussion}

We now interpret our key experimental results presented in Fig.~\ref{fig_5} with reference to the theoretical model presented in the previous section. For low Cr thicknesses ($d_\mathrm{Cr} =$ 1.3~nm), corresponding to antiparallel alignment of the neighboring NiFe and SFM interfaces, NiFe damping $\alpha$ is reduced compared to the reference NiFe film. This is likely due to a stronger spin-pumping contribution to the magnetic damping of the reference NiFe film grown onto a relatively thick two-layer buffer, Cr(5)/Cu(5), which should efficiently dissipate spin-pumped current regardless of the magnetic field applied~\cite{Tserkovnyak2005}. The other limit, that of the all-parallel alignment ($d_\mathrm{Cr} =$ 2.0, 5.0~nm) shows $\alpha$ larger than that for the reference NiFe film. This must be attributed to a larger $g_\mathrm{SFM}$ of SFM, indicating an enhanced dissipation of the spin-pumped current by the SFM in this magnetic state and, consequently, an increase in the overall $\alpha$ of the free layer. The difference in $\alpha$ between the antiparallel (thin Cr) and parallel (thick Cr) configurations is naturally assigned to the changes in $g_\mathrm{SFM}$, according to Eq.~(\ref{gb}), via the variation in the effective thickness $L_\mathrm{SFM}$ of the SFM. This conclusion follows also from the comparison of the two all-parallel cases, where $\alpha$ for $d_\mathrm{Cr} =$ 2.0~nm is smaller than that for $d_\mathrm{Cr} =$ 5.0~nm. However, the rather steep decrease in $\alpha$ upon relatively small changes in the Cr thickness, from $d_\mathrm{Cr} =$ 2.0~nm (parallel state) to $d_\mathrm{Cr} =$ 1.3~nm (antiparallel state), can be attributed to the spin-conductivity term in Eq.~(\ref{gb}): $\sigma_{\uparrow} \sigma_{\downarrow}/ (\sigma_{\uparrow} + \sigma_{\downarrow}) $. Here, the effective conductances $\sigma_{\uparrow,\downarrow}$ for the SFM depend upon the magnetic states of the Fe layers constituting the SFM. This factor is expected to be smaller in the anti-collinear configuration, similar to the physics underlying the giant magnetoresistance effect~\cite{Valet1993}.

The SFM enters a non-collinear regime for the intermediate Cr thicknesses, $d_\mathrm{Cr} =$ 1.5 and 1.7~nm, in which the spin-flip scattering is expected to be much stronger due to the increased magnetic inhomogeneity.  This leads to a much smaller $\lambda_{\mathrm{SFM}}$ and, hence, larger $g_{\mathrm{SFM}}$. We indeed observe a larger damping in the non-collinear configuration of the SFM (Fig.~\ref{fig_5}). In addition to the above outlined mechanism, spin-torque-like absorption of the transverse component of the pumped spin current at the N/SFM interface may also contribute to a larger spin absorption by the SFM layer in the non-collinear configuration. This latter absorption mechanism is not captured by the simplified model based on the longitudinal spin transport considered above.

Based on our observations, one can imagine a number of approaches for ex-situ control of magnetic damping in multilayers. In a broadband FMR configuration, for example, one can modulate the damping by varying the operating frequency. Lower frequencies would correspond to lower in-plane resonance fields, where the layers of interest can be antiparallel (lower dissipation). Higher frequencies require higher resonance fields, at which the layers can be saturated in the same direction (larger dissipation). Another approach would be to employ a synthetic ferrimagnet with temperature-dependent interlayer exchange, of type demonstrated in Refs.~\onlinecite{Polishchuk2017,Polishchuk2017a}. A relatively small change in temperature ($\approx$~50~K) can trigger the antiparallel-to-parallel transition in such SFM, which could be used to effectively tune the magnetic relaxation of the free layer.

\section{Conclusion}

We study intrinsic and extrinsic magnetic relaxation of a soft ferromagnetic layer interfaced with a synthetic ferrimagnet, using in-plane ferromagnetic resonance and magnetometry measurements combined with theoretical analysis. A parallel-to-antiparallel transition of the mutual orientation of the resonating free layer and its nearest SFM interface is achieved by controlling the strength of the indirect exchange coupling within the SFM. The associated dependence in the measured magnetic damping is non-monotonous: an intermediate non-collinear magnetic configuration in the structure results in a significantly enhanced spin relaxation in the system. This behavior is interpreted as arising from a varying effectiveness of the SFM in absorbing and dissipating the spin pumping current emitted by the free layer. In addition to an increased spin current absorption by a thicker SFM, our results highlight the roles of a stronger spin-flip in the non-collinear state as well as a modified effective conductance in the antiparallel state, akin to the giant magnetoresistance effect, of the SFM. Our results demonstrate the possibility of a high tunability of the magnetic damping in soft ferromagnets via relatively small changes in the magnetic multilayer layout.

 
Support from the Swedish Research Council (VR Grant No.~2014-4548) and the Swedish Stiftelse Olle Engkvist Byggm\"astare, Volkswagen Foundation (Grant No.~90418), the National Academy of Sciences of Ukraine (Project No.~0118U003265), the Research Council of Norway through its Centers of Excellence funding scheme (Project No.~262633, ``QuSpin''), the State Fund for Fundamental Research of Ukraine (Grant No.~F76/34-2018), the Department of Targeted Training of Taras Shevchenko National University of Kyiv at the National Academy of Sciences of Ukraine (Grant No.~0118U003837) are gratefully acknowledged.

\bibliography{References}

\begin{thebibliography}{28}%
\makeatletter
\providecommand \@ifxundefined [1]{%
 \@ifx{#1\undefined}
}%
\providecommand \@ifnum [1]{%
 \ifnum #1\expandafter \@firstoftwo
 \else \expandafter \@secondoftwo
 \fi
}%
\providecommand \@ifx [1]{%
 \ifx #1\expandafter \@firstoftwo
 \else \expandafter \@secondoftwo
 \fi
}%
\providecommand \natexlab [1]{#1}%
\providecommand \enquote  [1]{``#1''}%
\providecommand \bibnamefont  [1]{#1}%
\providecommand \bibfnamefont [1]{#1}%
\providecommand \citenamefont [1]{#1}%
\providecommand \href@noop [0]{\@secondoftwo}%
\providecommand \href [0]{\begingroup \@sanitize@url \@href}%
\providecommand \@href[1]{\@@startlink{#1}\@@href}%
\providecommand \@@href[1]{\endgroup#1\@@endlink}%
\providecommand \@sanitize@url [0]{\catcode `\\12\catcode `\$12\catcode
  `\&12\catcode `\#12\catcode `\^12\catcode `\_12\catcode `\%12\relax}%
\providecommand \@@startlink[1]{}%
\providecommand \@@endlink[0]{}%
\providecommand \url  [0]{\begingroup\@sanitize@url \@url }%
\providecommand \@url [1]{\endgroup\@href {#1}{\urlprefix }}%
\providecommand \urlprefix  [0]{URL }%
\providecommand \Eprint [0]{\href }%
\providecommand \doibase [0]{http://dx.doi.org/}%
\providecommand \selectlanguage [0]{\@gobble}%
\providecommand \bibinfo  [0]{\@secondoftwo}%
\providecommand \bibfield  [0]{\@secondoftwo}%
\providecommand \translation [1]{[#1]}%
\providecommand \BibitemOpen [0]{}%
\providecommand \bibitemStop [0]{}%
\providecommand \bibitemNoStop [0]{.\EOS\space}%
\providecommand \EOS [0]{\spacefactor3000\relax}%
\providecommand \BibitemShut  [1]{\csname bibitem#1\endcsname}%
\let\auto@bib@innerbib\@empty
\bibitem [{\citenamefont {Tserkovnyak}\ \emph
  {et~al.}(2002{\natexlab{a}})\citenamefont {Tserkovnyak}, \citenamefont
  {Brataas},\ and\ \citenamefont {Bauer}}]{Tserkovnyak2002}%
  \BibitemOpen
  \bibfield  {author} {\bibinfo {author} {\bibfnamefont {Y.}~\bibnamefont
  {Tserkovnyak}}, \bibinfo {author} {\bibfnamefont {A.}~\bibnamefont
  {Brataas}}, \ and\ \bibinfo {author} {\bibfnamefont {G.~E.~W.}\ \bibnamefont
  {Bauer}},\ }\href {\doibase 10.1103/physrevlett.88.117601} {\bibfield
  {journal} {\bibinfo  {journal} {Physical Review Letters}\ }\textbf {\bibinfo
  {volume} {88}},\ \bibinfo {pages} {117601} (\bibinfo {year}
  {2002}{\natexlab{a}})}\BibitemShut {NoStop}%
\bibitem [{\citenamefont {Brataas}\ \emph {et~al.}(2002)\citenamefont
  {Brataas}, \citenamefont {Tserkovnyak}, \citenamefont {Bauer},\ and\
  \citenamefont {Halperin}}]{Brataas2002}%
  \BibitemOpen
  \bibfield  {author} {\bibinfo {author} {\bibfnamefont {A.}~\bibnamefont
  {Brataas}}, \bibinfo {author} {\bibfnamefont {Y.}~\bibnamefont
  {Tserkovnyak}}, \bibinfo {author} {\bibfnamefont {G.~E.~W.}\ \bibnamefont
  {Bauer}}, \ and\ \bibinfo {author} {\bibfnamefont {B.~I.}\ \bibnamefont
  {Halperin}},\ }\href {\doibase 10.1103/physrevb.66.060404} {\bibfield
  {journal} {\bibinfo  {journal} {Physical Review B}\ }\textbf {\bibinfo
  {volume} {66}},\ \bibinfo {pages} {060404(R)} (\bibinfo {year}
  {2002})}\BibitemShut {NoStop}%
\bibitem [{\citenamefont {Ando}\ \emph {et~al.}(2011)\citenamefont {Ando},
  \citenamefont {Takahashi}, \citenamefont {Ieda}, \citenamefont {Kurebayashi},
  \citenamefont {Trypiniotis}, \citenamefont {Barnes}, \citenamefont
  {Maekawa},\ and\ \citenamefont {Saitoh}}]{Ando2011}%
  \BibitemOpen
  \bibfield  {author} {\bibinfo {author} {\bibfnamefont {K.}~\bibnamefont
  {Ando}}, \bibinfo {author} {\bibfnamefont {S.}~\bibnamefont {Takahashi}},
  \bibinfo {author} {\bibfnamefont {J.}~\bibnamefont {Ieda}}, \bibinfo {author}
  {\bibfnamefont {H.}~\bibnamefont {Kurebayashi}}, \bibinfo {author}
  {\bibfnamefont {T.}~\bibnamefont {Trypiniotis}}, \bibinfo {author}
  {\bibfnamefont {C.~H.~W.}\ \bibnamefont {Barnes}}, \bibinfo {author}
  {\bibfnamefont {S.}~\bibnamefont {Maekawa}}, \ and\ \bibinfo {author}
  {\bibfnamefont {E.}~\bibnamefont {Saitoh}},\ }\href {\doibase
  10.1038/nmat3052} {\bibfield  {journal} {\bibinfo  {journal} {Nature
  Materials}\ }\textbf {\bibinfo {volume} {10}},\ \bibinfo {pages} {655}
  (\bibinfo {year} {2011})}\BibitemShut {NoStop}%
\bibitem [{\citenamefont {Tserkovnyak}\ \emph {et~al.}(2005)\citenamefont
  {Tserkovnyak}, \citenamefont {Brataas}, \citenamefont {Bauer},\ and\
  \citenamefont {Halperin}}]{Tserkovnyak2005}%
  \BibitemOpen
  \bibfield  {author} {\bibinfo {author} {\bibfnamefont {Y.}~\bibnamefont
  {Tserkovnyak}}, \bibinfo {author} {\bibfnamefont {A.}~\bibnamefont
  {Brataas}}, \bibinfo {author} {\bibfnamefont {G.~E.~W.}\ \bibnamefont
  {Bauer}}, \ and\ \bibinfo {author} {\bibfnamefont {B.~I.}\ \bibnamefont
  {Halperin}},\ }\href {\doibase 10.1103/RevModPhys.77.1375} {\bibfield
  {journal} {\bibinfo  {journal} {Rev. Mod. Phys.}\ }\textbf {\bibinfo {volume}
  {77}},\ \bibinfo {pages} {1375} (\bibinfo {year} {2005})}\BibitemShut
  {NoStop}%
\bibitem [{\citenamefont {Brataas}\ \emph {et~al.}(2012)\citenamefont
  {Brataas}, \citenamefont {Tserkovnyak}, \citenamefont {Bauer},\ and\
  \citenamefont {Kelly}}]{SpinCurrent}%
  \BibitemOpen
  \bibfield  {author} {\bibinfo {author} {\bibfnamefont {A.}~\bibnamefont
  {Brataas}}, \bibinfo {author} {\bibfnamefont {Y.}~\bibnamefont
  {Tserkovnyak}}, \bibinfo {author} {\bibfnamefont {G.~E.~W.}\ \bibnamefont
  {Bauer}}, \ and\ \bibinfo {author} {\bibfnamefont {P.~J.}\ \bibnamefont
  {Kelly}},\ }\enquote {\bibinfo {title} {Spin current},}\ \ (\bibinfo {year}
  {2012})\ Chap.\ \bibinfo {chapter} {Spin pumping and spin transfer}, p.\
  \bibinfo {pages} {442}\BibitemShut {NoStop}%
\bibitem [{\citenamefont {Heinrich}\ \emph {et~al.}(2003)\citenamefont
  {Heinrich}, \citenamefont {Tserkovnyak}, \citenamefont {Woltersdorf},
  \citenamefont {Brataas}, \citenamefont {Urban},\ and\ \citenamefont
  {Bauer}}]{Heinrich2003}%
  \BibitemOpen
  \bibfield  {author} {\bibinfo {author} {\bibfnamefont {B.}~\bibnamefont
  {Heinrich}}, \bibinfo {author} {\bibfnamefont {Y.}~\bibnamefont
  {Tserkovnyak}}, \bibinfo {author} {\bibfnamefont {G.}~\bibnamefont
  {Woltersdorf}}, \bibinfo {author} {\bibfnamefont {A.}~\bibnamefont
  {Brataas}}, \bibinfo {author} {\bibfnamefont {R.}~\bibnamefont {Urban}}, \
  and\ \bibinfo {author} {\bibfnamefont {G.~E.~W.}\ \bibnamefont {Bauer}},\
  }\href {\doibase 10.1103/physrevlett.90.187601} {\bibfield  {journal}
  {\bibinfo  {journal} {Physical Review Letters}\ }\textbf {\bibinfo {volume}
  {90}},\ \bibinfo {pages} {187601} (\bibinfo {year} {2003})}\BibitemShut
  {NoStop}%
\bibitem [{\citenamefont {Brataas}\ \emph {et~al.}(2000)\citenamefont
  {Brataas}, \citenamefont {Nazarov},\ and\ \citenamefont
  {Bauer}}]{Brataas2000}%
  \BibitemOpen
  \bibfield  {author} {\bibinfo {author} {\bibfnamefont {A.}~\bibnamefont
  {Brataas}}, \bibinfo {author} {\bibfnamefont {Y.~V.}\ \bibnamefont
  {Nazarov}}, \ and\ \bibinfo {author} {\bibfnamefont {G.~E.~W.}\ \bibnamefont
  {Bauer}},\ }\href {\doibase 10.1103/physrevlett.84.2481} {\bibfield
  {journal} {\bibinfo  {journal} {Physical Review Letters}\ }\textbf {\bibinfo
  {volume} {84}},\ \bibinfo {pages} {2481} (\bibinfo {year}
  {2000})}\BibitemShut {NoStop}%
\bibitem [{\citenamefont {Tserkovnyak}\ \emph
  {et~al.}(2002{\natexlab{b}})\citenamefont {Tserkovnyak}, \citenamefont
  {Brataas},\ and\ \citenamefont {Bauer}}]{Tserkovnyak2002a}%
  \BibitemOpen
  \bibfield  {author} {\bibinfo {author} {\bibfnamefont {Y.}~\bibnamefont
  {Tserkovnyak}}, \bibinfo {author} {\bibfnamefont {A.}~\bibnamefont
  {Brataas}}, \ and\ \bibinfo {author} {\bibfnamefont {G.~E.~W.}\ \bibnamefont
  {Bauer}},\ }\href {\doibase 10.1103/physrevb.66.224403} {\bibfield  {journal}
  {\bibinfo  {journal} {Physical Review B}\ }\textbf {\bibinfo {volume} {66}},\
  \bibinfo {pages} {224403} (\bibinfo {year} {2002}{\natexlab{b}})}\BibitemShut
  {NoStop}%
\bibitem [{\citenamefont {Ghosh}\ \emph {et~al.}(2012)\citenamefont {Ghosh},
  \citenamefont {Auffret}, \citenamefont {Ebels},\ and\ \citenamefont
  {Bailey}}]{Ghosh2012}%
  \BibitemOpen
  \bibfield  {author} {\bibinfo {author} {\bibfnamefont {A.}~\bibnamefont
  {Ghosh}}, \bibinfo {author} {\bibfnamefont {S.}~\bibnamefont {Auffret}},
  \bibinfo {author} {\bibfnamefont {U.}~\bibnamefont {Ebels}}, \ and\ \bibinfo
  {author} {\bibfnamefont {W.~E.}\ \bibnamefont {Bailey}},\ }\href {\doibase
  10.1103/physrevlett.109.127202} {\bibfield  {journal} {\bibinfo  {journal}
  {Physical Review Letters}\ }\textbf {\bibinfo {volume} {109}},\ \bibinfo
  {pages} {127202} (\bibinfo {year} {2012})}\BibitemShut {NoStop}%
\bibitem [{\citenamefont {Merodio}\ \emph {et~al.}(2014)\citenamefont
  {Merodio}, \citenamefont {Ghosh}, \citenamefont {Lemonias}, \citenamefont
  {Gautier}, \citenamefont {Ebels}, \citenamefont {Chshiev}, \citenamefont
  {B{\'{e}}a}, \citenamefont {Baltz},\ and\ \citenamefont
  {Bailey}}]{Merodio2014}%
  \BibitemOpen
  \bibfield  {author} {\bibinfo {author} {\bibfnamefont {P.}~\bibnamefont
  {Merodio}}, \bibinfo {author} {\bibfnamefont {A.}~\bibnamefont {Ghosh}},
  \bibinfo {author} {\bibfnamefont {C.}~\bibnamefont {Lemonias}}, \bibinfo
  {author} {\bibfnamefont {E.}~\bibnamefont {Gautier}}, \bibinfo {author}
  {\bibfnamefont {U.}~\bibnamefont {Ebels}}, \bibinfo {author} {\bibfnamefont
  {M.}~\bibnamefont {Chshiev}}, \bibinfo {author} {\bibfnamefont
  {H.}~\bibnamefont {B{\'{e}}a}}, \bibinfo {author} {\bibfnamefont
  {V.}~\bibnamefont {Baltz}}, \ and\ \bibinfo {author} {\bibfnamefont {W.~E.}\
  \bibnamefont {Bailey}},\ }\href {\doibase 10.1063/1.4862971} {\bibfield
  {journal} {\bibinfo  {journal} {Applied Physics Letters}\ }\textbf {\bibinfo
  {volume} {104}},\ \bibinfo {pages} {032406} (\bibinfo {year}
  {2014})}\BibitemShut {NoStop}%
\bibitem [{\citenamefont {Mizukami}\ \emph {et~al.}(2001)\citenamefont
  {Mizukami}, \citenamefont {Ando},\ and\ \citenamefont
  {Miyazaki}}]{Mizukami2001}%
  \BibitemOpen
  \bibfield  {author} {\bibinfo {author} {\bibfnamefont {S.}~\bibnamefont
  {Mizukami}}, \bibinfo {author} {\bibfnamefont {Y.}~\bibnamefont {Ando}}, \
  and\ \bibinfo {author} {\bibfnamefont {T.}~\bibnamefont {Miyazaki}},\ }\href
  {\doibase 10.1143/jjap.40.580} {\bibfield  {journal} {\bibinfo  {journal}
  {Japanese Journal of Applied Physics}\ }\textbf {\bibinfo {volume} {40}},\
  \bibinfo {pages} {580} (\bibinfo {year} {2001})}\BibitemShut {NoStop}%
\bibitem [{\citenamefont {Foros}\ \emph {et~al.}(2005)\citenamefont {Foros},
  \citenamefont {Woltersdorf}, \citenamefont {Heinrich},\ and\ \citenamefont
  {Brataas}}]{Foros2005}%
  \BibitemOpen
  \bibfield  {author} {\bibinfo {author} {\bibfnamefont {J.}~\bibnamefont
  {Foros}}, \bibinfo {author} {\bibfnamefont {G.}~\bibnamefont {Woltersdorf}},
  \bibinfo {author} {\bibfnamefont {B.}~\bibnamefont {Heinrich}}, \ and\
  \bibinfo {author} {\bibfnamefont {A.}~\bibnamefont {Brataas}},\ }\href
  {\doibase 10.1063/1.1853131} {\bibfield  {journal} {\bibinfo  {journal}
  {Journal of Applied Physics}\ }\textbf {\bibinfo {volume} {97}},\ \bibinfo
  {pages} {10A714} (\bibinfo {year} {2005})}\BibitemShut {NoStop}%
\bibitem [{\citenamefont {Jungwirth}\ \emph {et~al.}(2016)\citenamefont
  {Jungwirth}, \citenamefont {Marti}, \citenamefont {Wadley},\ and\
  \citenamefont {Wunderlich}}]{Jungwirth2016}%
  \BibitemOpen
  \bibfield  {author} {\bibinfo {author} {\bibfnamefont {T.}~\bibnamefont
  {Jungwirth}}, \bibinfo {author} {\bibfnamefont {X.}~\bibnamefont {Marti}},
  \bibinfo {author} {\bibfnamefont {P.}~\bibnamefont {Wadley}}, \ and\ \bibinfo
  {author} {\bibfnamefont {J.}~\bibnamefont {Wunderlich}},\ }\href {\doibase
  10.1038/nnano.2016.18} {\bibfield  {journal} {\bibinfo  {journal} {Nature
  Nanotechnology}\ }\textbf {\bibinfo {volume} {11}},\ \bibinfo {pages} {231}
  (\bibinfo {year} {2016})}\BibitemShut {NoStop}%
\bibitem [{\citenamefont {Bass}\ and\ \citenamefont {Pratt}(2007)}]{Bass2007}%
  \BibitemOpen
  \bibfield  {author} {\bibinfo {author} {\bibfnamefont {J.}~\bibnamefont
  {Bass}}\ and\ \bibinfo {author} {\bibfnamefont {W.~P.}\ \bibnamefont
  {Pratt}},\ }\href {\doibase 10.1088/0953-8984/19/18/183201} {\bibfield
  {journal} {\bibinfo  {journal} {Journal of Physics: Condensed Matter}\
  }\textbf {\bibinfo {volume} {19}},\ \bibinfo {pages} {183201} (\bibinfo
  {year} {2007})}\BibitemShut {NoStop}%
\bibitem [{\citenamefont {Slonczewski}(1996)}]{Slonczewski1996}%
  \BibitemOpen
  \bibfield  {author} {\bibinfo {author} {\bibfnamefont {J.}~\bibnamefont
  {Slonczewski}},\ }\href {\doibase 10.1016/0304-8853(96)00062-5} {\bibfield
  {journal} {\bibinfo  {journal} {Journal of Magnetism and Magnetic Materials}\
  }\textbf {\bibinfo {volume} {159}},\ \bibinfo {pages} {L1} (\bibinfo {year}
  {1996})}\BibitemShut {NoStop}%
\bibitem [{\citenamefont {Taniguchi}\ and\ \citenamefont
  {Imamura}(2007)}]{Taniguchi2007}%
  \BibitemOpen
  \bibfield  {author} {\bibinfo {author} {\bibfnamefont {T.}~\bibnamefont
  {Taniguchi}}\ and\ \bibinfo {author} {\bibfnamefont {H.}~\bibnamefont
  {Imamura}},\ }\href {\doibase 10.1103/physrevb.76.092402} {\bibfield
  {journal} {\bibinfo  {journal} {Physical Review B}\ }\textbf {\bibinfo
  {volume} {76}},\ \bibinfo {pages} {092402} (\bibinfo {year}
  {2007})}\BibitemShut {NoStop}%
\bibitem [{\citenamefont {Chiba}\ \emph {et~al.}(2015)\citenamefont {Chiba},
  \citenamefont {Bauer},\ and\ \citenamefont {Takahashi}}]{Chiba2015}%
  \BibitemOpen
  \bibfield  {author} {\bibinfo {author} {\bibfnamefont {T.}~\bibnamefont
  {Chiba}}, \bibinfo {author} {\bibfnamefont {G.~E.~W.}\ \bibnamefont {Bauer}},
  \ and\ \bibinfo {author} {\bibfnamefont {S.}~\bibnamefont {Takahashi}},\
  }\href {\doibase 10.1103/physrevb.92.054407} {\bibfield  {journal} {\bibinfo
  {journal} {Physical Review B}\ }\textbf {\bibinfo {volume} {92}},\ \bibinfo
  {pages} {054407} (\bibinfo {year} {2015})}\BibitemShut {NoStop}%
\bibitem [{\citenamefont {Mosendz}\ \emph {et~al.}(2008)\citenamefont
  {Mosendz}, \citenamefont {Kardasz},\ and\ \citenamefont
  {Heinrich}}]{Mosendz2008}%
  \BibitemOpen
  \bibfield  {author} {\bibinfo {author} {\bibfnamefont {O.}~\bibnamefont
  {Mosendz}}, \bibinfo {author} {\bibfnamefont {B.}~\bibnamefont {Kardasz}}, \
  and\ \bibinfo {author} {\bibfnamefont {B.}~\bibnamefont {Heinrich}},\ }\href
  {\doibase 10.1063/1.2830645} {\bibfield  {journal} {\bibinfo  {journal}
  {Journal of Applied Physics}\ }\textbf {\bibinfo {volume} {103}},\ \bibinfo
  {pages} {07B505} (\bibinfo {year} {2008})}\BibitemShut {NoStop}%
\bibitem [{\citenamefont {Demirtas}\ \emph {et~al.}(2011)\citenamefont
  {Demirtas}, \citenamefont {Salamon},\ and\ \citenamefont
  {Koymen}}]{Demirtas2011}%
  \BibitemOpen
  \bibfield  {author} {\bibinfo {author} {\bibfnamefont {S.}~\bibnamefont
  {Demirtas}}, \bibinfo {author} {\bibfnamefont {M.~B.}\ \bibnamefont
  {Salamon}}, \ and\ \bibinfo {author} {\bibfnamefont {A.~R.}\ \bibnamefont
  {Koymen}},\ }\href {\doibase 10.1063/1.3592298} {\bibfield  {journal}
  {\bibinfo  {journal} {Journal of Applied Physics}\ }\textbf {\bibinfo
  {volume} {109}},\ \bibinfo {pages} {113919} (\bibinfo {year}
  {2011})}\BibitemShut {NoStop}%
\bibitem [{\citenamefont {Nogu{\'{e}}s}\ \emph {et~al.}(2005)\citenamefont
  {Nogu{\'{e}}s}, \citenamefont {Sort}, \citenamefont {Langlais}, \citenamefont
  {Skumryev}, \citenamefont {Suri{\~{n}}ach}, \citenamefont {Mu{\~{n}}oz},\
  and\ \citenamefont {Bar{\'{o}}}}]{Nogues2005}%
  \BibitemOpen
  \bibfield  {author} {\bibinfo {author} {\bibfnamefont {J.}~\bibnamefont
  {Nogu{\'{e}}s}}, \bibinfo {author} {\bibfnamefont {J.}~\bibnamefont {Sort}},
  \bibinfo {author} {\bibfnamefont {V.}~\bibnamefont {Langlais}}, \bibinfo
  {author} {\bibfnamefont {V.}~\bibnamefont {Skumryev}}, \bibinfo {author}
  {\bibfnamefont {S.}~\bibnamefont {Suri{\~{n}}ach}}, \bibinfo {author}
  {\bibfnamefont {J.}~\bibnamefont {Mu{\~{n}}oz}}, \ and\ \bibinfo {author}
  {\bibfnamefont {M.}~\bibnamefont {Bar{\'{o}}}},\ }\href {\doibase
  10.1016/j.physrep.2005.08.004} {\bibfield  {journal} {\bibinfo  {journal}
  {Physics Reports}\ }\textbf {\bibinfo {volume} {422}},\ \bibinfo {pages} {65}
  (\bibinfo {year} {2005})}\BibitemShut {NoStop}%
\bibitem [{\citenamefont {Timopheev}\ \emph {et~al.}(2014)\citenamefont
  {Timopheev}, \citenamefont {Pogorelov}, \citenamefont {Cardoso},
  \citenamefont {Freitas}, \citenamefont {Kakazei},\ and\ \citenamefont
  {Sobolev}}]{Timopheev2014}%
  \BibitemOpen
  \bibfield  {author} {\bibinfo {author} {\bibfnamefont {A.~A.}\ \bibnamefont
  {Timopheev}}, \bibinfo {author} {\bibfnamefont {Y.~G.}\ \bibnamefont
  {Pogorelov}}, \bibinfo {author} {\bibfnamefont {S.}~\bibnamefont {Cardoso}},
  \bibinfo {author} {\bibfnamefont {P.~P.}\ \bibnamefont {Freitas}}, \bibinfo
  {author} {\bibfnamefont {G.~N.}\ \bibnamefont {Kakazei}}, \ and\ \bibinfo
  {author} {\bibfnamefont {N.~A.}\ \bibnamefont {Sobolev}},\ }\href {\doibase
  10.1103/physrevb.89.144410} {\bibfield  {journal} {\bibinfo  {journal}
  {Physical Review B}\ }\textbf {\bibinfo {volume} {89}},\ \bibinfo {pages}
  {144410} (\bibinfo {year} {2014})}\BibitemShut {NoStop}%
\bibitem [{\citenamefont {Kravets}\ \emph {et~al.}(2016)\citenamefont
  {Kravets}, \citenamefont {Polishchuk}, \citenamefont {Dzhezherya},
  \citenamefont {Tovstolytkin}, \citenamefont {Golub},\ and\ \citenamefont
  {Korenivski}}]{Kravets2016}%
  \BibitemOpen
  \bibfield  {author} {\bibinfo {author} {\bibfnamefont {A.~F.}\ \bibnamefont
  {Kravets}}, \bibinfo {author} {\bibfnamefont {D.~M.}\ \bibnamefont
  {Polishchuk}}, \bibinfo {author} {\bibfnamefont {Y.~I.}\ \bibnamefont
  {Dzhezherya}}, \bibinfo {author} {\bibfnamefont {A.~I.}\ \bibnamefont
  {Tovstolytkin}}, \bibinfo {author} {\bibfnamefont {V.~O.}\ \bibnamefont
  {Golub}}, \ and\ \bibinfo {author} {\bibfnamefont {V.}~\bibnamefont
  {Korenivski}},\ }\href {\doibase 10.1103/physrevb.94.064429} {\bibfield
  {journal} {\bibinfo  {journal} {Physical Review B}\ }\textbf {\bibinfo
  {volume} {94}},\ \bibinfo {pages} {064429} (\bibinfo {year}
  {2016})}\BibitemShut {NoStop}%
\bibitem [{\citenamefont {Grünberg}\ \emph {et~al.}(1987)\citenamefont
  {Grünberg}, \citenamefont {Schreiber}, \citenamefont {Pang}, \citenamefont
  {Walz}, \citenamefont {Brodsky},\ and\ \citenamefont
  {Sowers}}]{Gruenberg1987}%
  \BibitemOpen
  \bibfield  {author} {\bibinfo {author} {\bibfnamefont {P.}~\bibnamefont
  {Grünberg}}, \bibinfo {author} {\bibfnamefont {R.}~\bibnamefont
  {Schreiber}}, \bibinfo {author} {\bibfnamefont {Y.}~\bibnamefont {Pang}},
  \bibinfo {author} {\bibfnamefont {U.}~\bibnamefont {Walz}}, \bibinfo {author}
  {\bibfnamefont {M.~B.}\ \bibnamefont {Brodsky}}, \ and\ \bibinfo {author}
  {\bibfnamefont {H.}~\bibnamefont {Sowers}},\ }\href {\doibase
  10.1063/1.338656} {\bibfield  {journal} {\bibinfo  {journal} {Journal of
  Applied Physics}\ }\textbf {\bibinfo {volume} {61}},\ \bibinfo {pages} {3750}
  (\bibinfo {year} {1987})}\BibitemShut {NoStop}%
\bibitem [{\citenamefont {Polishchuk}\ \emph
  {et~al.}(2017{\natexlab{a}})\citenamefont {Polishchuk}, \citenamefont
  {Tykhonenko-Polishchuk}, \citenamefont {Kravets},\ and\ \citenamefont
  {Korenivski}}]{Polishchuk2017}%
  \BibitemOpen
  \bibfield  {author} {\bibinfo {author} {\bibfnamefont {D.~M.}\ \bibnamefont
  {Polishchuk}}, \bibinfo {author} {\bibfnamefont {Y.~O.}\ \bibnamefont
  {Tykhonenko-Polishchuk}}, \bibinfo {author} {\bibfnamefont {A.~F.}\
  \bibnamefont {Kravets}}, \ and\ \bibinfo {author} {\bibfnamefont
  {V.}~\bibnamefont {Korenivski}},\ }\href {\doibase
  10.1209/0295-5075/118/37006} {\bibfield  {journal} {\bibinfo  {journal}
  {{EPL} (Europhysics Letters)}\ }\textbf {\bibinfo {volume} {118}},\ \bibinfo
  {pages} {37006} (\bibinfo {year} {2017}{\natexlab{a}})}\BibitemShut {NoStop}%
\bibitem [{\citenamefont {Zhang}\ \emph {et~al.}(1994)\citenamefont {Zhang},
  \citenamefont {Zhou}, \citenamefont {Wigen},\ and\ \citenamefont
  {Ounadjela}}]{Zhang1994}%
  \BibitemOpen
  \bibfield  {author} {\bibinfo {author} {\bibfnamefont {Z.}~\bibnamefont
  {Zhang}}, \bibinfo {author} {\bibfnamefont {L.}~\bibnamefont {Zhou}},
  \bibinfo {author} {\bibfnamefont {P.~E.}\ \bibnamefont {Wigen}}, \ and\
  \bibinfo {author} {\bibfnamefont {K.}~\bibnamefont {Ounadjela}},\ }\href
  {\doibase 10.1103/physrevb.50.6094} {\bibfield  {journal} {\bibinfo
  {journal} {Physical Review B}\ }\textbf {\bibinfo {volume} {50}},\ \bibinfo
  {pages} {6094} (\bibinfo {year} {1994})}\BibitemShut {NoStop}%
\bibitem [{\citenamefont {Valet}\ and\ \citenamefont {Fert}(1993)}]{Valet1993}%
  \BibitemOpen
  \bibfield  {author} {\bibinfo {author} {\bibfnamefont {T.}~\bibnamefont
  {Valet}}\ and\ \bibinfo {author} {\bibfnamefont {A.}~\bibnamefont {Fert}},\
  }\href {\doibase 10.1103/physrevb.48.7099} {\bibfield  {journal} {\bibinfo
  {journal} {Physical Review B}\ }\textbf {\bibinfo {volume} {48}},\ \bibinfo
  {pages} {7099} (\bibinfo {year} {1993})}\BibitemShut {NoStop}%
\bibitem [{\citenamefont {Fabian}\ \emph {et~al.}(2010)\citenamefont {Fabian},
  \citenamefont {Matos-Abiague}, \citenamefont {Ertler}, \citenamefont
  {Stano},\ and\ \citenamefont {Žutić}}]{Fabian2010}%
  \BibitemOpen
  \bibfield  {author} {\bibinfo {author} {\bibfnamefont {J.}~\bibnamefont
  {Fabian}}, \bibinfo {author} {\bibfnamefont {A.}~\bibnamefont
  {Matos-Abiague}}, \bibinfo {author} {\bibfnamefont {C.}~\bibnamefont
  {Ertler}}, \bibinfo {author} {\bibfnamefont {P.}~\bibnamefont {Stano}}, \
  and\ \bibinfo {author} {\bibfnamefont {I.}~\bibnamefont {Žutić}},\ }\href
  {\doibase 10.2478/v10155-010-0086-8} {\bibfield  {journal} {\bibinfo
  {journal} {Acta Physica Slovaca}\ }\textbf {\bibinfo {volume} {57}},\
  \bibinfo {pages} {565} (\bibinfo {year} {2010})}\BibitemShut {NoStop}%
\bibitem [{\citenamefont {Polishchuk}\ \emph
  {et~al.}(2017{\natexlab{b}})\citenamefont {Polishchuk}, \citenamefont
  {Tykhonenko-Polishchuk}, \citenamefont {Holmgren}, \citenamefont {Kravets},\
  and\ \citenamefont {Korenivski}}]{Polishchuk2017a}%
  \BibitemOpen
  \bibfield  {author} {\bibinfo {author} {\bibfnamefont {D.~M.}\ \bibnamefont
  {Polishchuk}}, \bibinfo {author} {\bibfnamefont {Y.~O.}\ \bibnamefont
  {Tykhonenko-Polishchuk}}, \bibinfo {author} {\bibfnamefont {E.}~\bibnamefont
  {Holmgren}}, \bibinfo {author} {\bibfnamefont {A.~F.}\ \bibnamefont
  {Kravets}}, \ and\ \bibinfo {author} {\bibfnamefont {V.}~\bibnamefont
  {Korenivski}},\ }\href {\doibase 10.1103/physrevb.96.104427} {\bibfield
  {journal} {\bibinfo  {journal} {Physical Review B}\ }\textbf {\bibinfo
  {volume} {96}},\ \bibinfo {pages} {104427} (\bibinfo {year}
  {2017}{\natexlab{b}})}\BibitemShut {NoStop}%
\end{thebibliography}%

\end{document}